\documentclass[tightenlines,aps,floatfix,nofootinbib,11pt]{revtex4}
\usepackage{bm}

\newcommand{\be}{\begin{equation}}
\newcommand{\ee}{\end{equation}}
\newcommand{\ba}{\begin{eqnarray}}
\newcommand{\ea}{\end{eqnarray}}
\newcommand{\Li}{{\rm Li}}
\newcommand{\e}{{\epsilon}}

\begin{document}


\begin{titlepage}

\begin{flushright}
\vbox{
\begin{tabular}{l}
 \textsf{FERMILAB-PUB-07-060-T}\\
 \textsf{UH-511-1103-07}
\end{tabular}
}
\end{flushright}

\title{Two-loop QED corrections to Bhabha scattering}

\author{Thomas Becher\thanks{e-mail: becher@fnal.gov}} 
\affiliation{
Fermi National Accelerator Laboratory, P.O. Box 500, Batavia, IL 60510, USA
}
\email{becher@fnal.gov}
\author{Kirill Melnikov \thanks{
e-mail:  kirill@phys.hawaii.edu}}
\affiliation{Department of Physics and Astronomy, 
University of Hawaii, Honolulu, HI 96822, USA}
\email{kirill@phys.hawaii.edu}

\begin{abstract}
We obtain a simple relation between massless and massive 
scattering amplitudes in gauge theories in the limit where all kinematic invariants are large compared to particle masses. We use this relation to derive the  two-loop  QED corrections to large-angle Bhabha 
scattering. 
\end{abstract}

\maketitle

\thispagestyle{empty}
\end{titlepage}

\section{Introduction}

Bhabha scattering,  $e^+e^- \to e^+e^-$, is one of the basic processes 
at electron-positron colliders. It has sufficiently large cross-section 
to be employed as a reference process for  collider
luminosity measurements. 
To determine the luminosity, 
one takes the ratio of the number of electron-positron pairs observed in the detector and the cross-section for  Bhabha scattering computed theoretically. It follows that the accuracy of the luminosity determination is fully correlated with the precision of the theoretical description of Bhabha scattering. 

There are two kinematic regimes relevant for  Bhabha scattering. 
 {\it Small angle} scattering, $\theta \lesssim {\rm few~degrees}$, 
was employed at LEP and  SLC for the luminosity determination, while  {\it large angle} Bhabha scattering, $\theta \gtrsim 10~{\rm degrees}$, is used at flavor factories (BABAR, BELLE, BEPC-BES, CLEO-C, DAPHNE, VEPP-2M) for that purpose \cite{calame}. Large angle Bhabha scattering can also be used to determine the luminosity spectrum \cite{miller} at the International Linear Collider (ILC).

Because of the importance of  Bhabha scattering,  
the theoretical description of this process is quite advanced. In particular, 
several Monte Carlo event generators have been developed to describe the 
Bhabha scattering process \cite{mc}; as a rule, these programs
correctly reproduce the 
${\cal O}(\alpha)$ QED corrections and some parts of  higher order 
corrections enhanced by large logarithms. Further progress in the theoretical description of Bhabha scattering 
requires the computation of ${\cal O}(\alpha^2)$
next-to-next-to-leading (NNLO) QED corrections and the incorporation of 
those corrections  into Monte Carlo event generators. The virtual two-loop corrections to 
$e^+e^- \to e^+e^-$ scattering amplitude were calculated  in 
Ref.~\cite{dixon}  in the approximation where lepton masses were 
set to zero.  Unfortunately, this result is insufficient for the description 
of Bhabha scattering at flavor factories and the ILC where isolated 
leptons rather than ``QED jets'' can be observed. Furthermore, all existing event generators work with massive electrons. To include the NNLO results into these codes, it is thus necessary to keep 
the electron mass also in  fixed order calculations.

For all practical applications, having the logarithmic electron mass dependence of the NNLO corrections is sufficient. Corrections suppressed by powers of the electron mass are negligible, even at the smallest measured scattering angles. When performing the calculation of the cross section with a nonzero electron mass, one can thus  expand the relevant diagrams in powers of the electron mass. Even so, the evaluation of the two-loop corrections remains a formidable task. Results for some of the necessary loop integrals were presented in Refs.~\cite{Smirnov:2001cm,Bonciani:2003te, Davydychev:2003mv, Bonciani:2003cj, Heinrich:2004iq, Czakon:2004wm,Czakon:2006pa}, but so far only the part of the cross section which involves closed electron loops has been evaluated \cite{Bonciani:2004gi, Bonciani:2004qt,fer}.

It is possible to avoid the evaluation of the diagrams in the massive case. Instead, one can use the massless result of Ref.~\cite{dixon} and restore the mass dependence in the logarithmic approximation. For corrections 
that do not involve closed fermion loops, to which we will refer to as ``photonic'',
this was shown in Ref.~\cite{penin} where the phenomenologically relevant 
result for these $O(\alpha^2)$ corrections to the large-angle Bhabha scattering was first reported. However, the approach of Ref.~\cite{penin} is somewhat complicated since it requires to transform infrared and collinear $1/\epsilon$ poles, inherent to massless amplitudes computed in dimensional regularization, to $\ln\lambda$ and $\ln m _e$ terms in the massive amplitude where the photon mass $\lambda$ regularizes   
infrared divergences and the electron mass $m_e$ regularizes
collinear divergences. 

In this paper we point out that a much simpler procedure for deriving massive amplitudes from massless ones exists since the two amplitudes are related by multiplicative renormalization factors. These renormalization factors 
can be  deduced from the knowledge of massive and massless
electron Dirac  form-factors. The massive  amplitude constructed 
along these lines has its infrared 
divergences regularized dimensionally and collinear divergences 
regularized  by the electron mass. Beyond its simplicity, 
the advantage of this approach is that it can be directly applied to QCD whereas the method of Ref.~\cite{penin} relies on the photon mass as infrared regulator. In a recent paper, Mitov and Moch have obtained a similar relation between massless and massive amplitudes \cite{mitov}. In fact, for photonic corrections our relation reduces to their result. In addition, our method allows us to also treat contributions involving massive fermion loops.

We apply our method to compute the NNLO QED corrections to large-angle Bhabha scattering; we include both photonic 
corrections and contributions from closed lepton loops.
This 
calculation is a nontrivial application of our method.
In addition, it  provides an 
independent check of the computations of 
Refs.~\cite{penin,Bonciani:2004gi, Bonciani:2004qt,fer} with 
which we find  complete agreement\footnote{Note that Appendix B of Ref.~\cite{Bonciani:2004qt} 
contains a misprint \cite{ferpriv}.}. 
We also derive the  NNLO contribution from 
loops with leptons heavier than the electron, i.e.\ muons and tau leptons, which was not available in the literature.

The  paper is organized as follows. In the next Section we present our 
notation and discuss the perturbative expansion of the large-angle 
Bhabha scattering cross-section. In Section~\ref{mass} we explain the 
factorization formula considering the electron Dirac 
form factor as an example.
In Section~\ref{comp} we apply the factorization formula to compute the 
NNLO QED corrections to Bhabha scattering. 
 We conclude in Section~\ref{conc}. 
Some useful formulas are collected in the Appendix.

\section{Notation}
\label{not}

Consider the process $e^+(p_1) + e^-(p_2) \to e^+(p_3) + e^-(p_4)$
for energies and  scattering angles such that  the absolute values
of all  kinematic invariants
$(p_1+p_2)^2 =s$, $(p_1-p_3)^2 = t$ and $(p_1-p_4)^2 = u$ are 
much larger than the electron mass  squared, $ s,|t|,|u| \gg m_e^2$.

We compute the Bhabha scattering cross-section perturbatively in the on-shell scheme where
 the fine structure constant $\alpha$ is defined through the photon propagator 
at zero momentum transfer. Neglecting corrections suppressed by powers 
of the electron mass, we write
\be
\frac{{\rm d} \sigma}{{\rm d} \Omega} 
= \frac{\alpha^2}{s} \left ( \frac{1-x+x^2}{x} \right )^2 \left [
1 + \left ( \frac{\alpha}{\pi} \right ) \delta_1
+  \left ( \frac{\alpha}{\pi} \right )^2 \delta_2 + {\cal O}(\alpha^3)
\right ],
\label{eq1}
\ee
where $x = (1-\cos \theta)/2$ and $\theta$ 
is the scattering angle. We take into account 
electron and muon contributions to photon vacuum polarization diagrams in the approximation $s \gg m_\mu^2 \gg m_e^2$. For phenomenological 
applications, the contribution of $\tau$ leptons may be required; 
it can be obtained from the formulas below with the obvious modification 
$m_\mu \to m_\tau$ provided that the 
high-energy condition $s \gg m_\tau^2$ is valid.

The higher order corrections to the Bhabha scattering cross-section 
depend logarithmically 
on the mass of the electron. Also, in order to arrive at a physically 
meaningful result, we need to allow for soft radiation with the energy 
of each emitted photon below some value $\omega_{\rm cut} \ll m$. 
The perturbative corrections depend logarithmically 
on $\omega_{\rm cut}$. The corrections sensitive to soft and collinear
physics are numerically
enhanced relative to other corrections; it is therefore customary to 
separate out those corrections when presenting results for perturbative 
coefficients. The ${\cal O}(\alpha)$ correction in Eq.~(\ref{eq1}) is 
well-known \cite{one-loop}; 
in the limit of small electron mass it can be written 
as
\be
\delta_1 = \left ( 4L_{\rm soft} + 3 
+ \frac{2}{3} N_f
\right )  
\ln \left ( \frac{s}{m_e^2} \right ) + \delta_1^{(0)},
\label{eq3}
\ee
where $L_{\rm soft} = \ln(2 \omega_{\rm cut}/\sqrt{s})$. We have introduced 
a label $N_f$ to distinguish corrections due to closed lepton loops. 
In particular, including electron and muon loops corresponds to $N_f = N_e + N_\mu$ where $N_i$ is the number of leptons of the $i$th flavor. To obtain numerical results 
one has to set 
$N_e = 1$ and $N_\mu = 1$.

The part of the one-loop correction that is not enhanced by the 
logarithm of the electron mass to center-of-mass energy ratio reads
\ba
\delta_1^{(0)}&& = \left [-4 + 4 \ln \frac{x}{1-x}  \right ]L_{\rm soft}
- \frac{2N_\mu}{3} \ln \frac{m_\mu^2}{m_e^2}
-4 - \frac{2\pi^2}{3} 
\nonumber \\
&& 
-2 \Li_2(x) + 2\Li_2(1-x) -\frac{10N_f}{9} + f(x).
\ea
The function $f(x)$ is defined as
\ba
&& f(x)  = (1-x+x^2)^{-2} \left \{ 
\left ( \frac{1}{3} - \frac{2}{3}x + \frac{9}{4}x^2-\frac{13}{6}x^3
+ \frac{4}{3}x^4 \right )\pi^2 
+ \left ( 3 -4x + \frac{9}{2}x^2 - \frac{3}{2}x^3 \right ) \ln(x)
\right.
\nonumber \\
&& \left. 
+ \left ( \frac{3}{4}x - \frac{x^2}{4} - \frac{3}{4}x^3 + x^4 \right )\ln^2(x)
+ \left [ -\frac{1}{2}x - \frac{1}{2}x^3
+ \left (2 -4x + \frac{7}{2}x^2 - x^3 \right) \ln(x) \right ] \ln(1-x)
\right.
\nonumber \\
&& \left.
+ \left (
-1+\frac{5}{2}x - \frac{7}{2}x^2 + \frac{5}{2}x^3-x^4 \right ) \ln^2(1-x)
+ N_f \left ( \frac{2}{3}-\frac{x}{3} \right )(1-x+x^2)\ln(x)
 \right \}.
\label{eq5}
\ea

The second term in the perturbative expansion of 
the Bhabha scattering cross-section is enhanced by up to three powers 
of the  logarithm of the electron mass.  
We write the NNLO correction as 
\be
\delta_2 = -\frac{N_f}{9}\ln^3 \left ( \frac{s}{m_e^2} \right )
+ \delta_2^{(2)} \ln^2 \left ( \frac{s}{m_e^2} \right ) 
+ \delta_2^{(1)} \ln \left ( \frac{s}{m_e^2} \right ) + \delta_2^{(0)},
\label{eq6}
\ee
where
\ba
&&  \delta_2^{(2)} = 8 L_{\rm soft}^2 + \left ( 12 + \frac{8}{3} N_f \right )
L_{\rm soft} + \frac{9}{2}
 + N_f \left ( -
\frac{1}{3} \ln \frac{x}{1-x} + \frac{55}{18}
\right ) 
+ \frac{N_\mu}{3} \ln \frac{m_\mu^2}{m_e^2}
+ \frac{N_f^2}{3}; 
\\
&&  \delta_2^{(1)}= \left[ -16 + 16 \ln \frac{x}{1-x} \right ] L_{\rm soft}^2
+ \left [-28 - \frac{8}{3}\pi^2 
+ 12 \ln \frac{x}{1-x} - 8 \Li_2(x)
+ 8 \Li_2(1-x) 
\right. 
 \\
&& \left.
+ N_f \left (\frac{8}{3} \ln \frac{x}{1-x} - \frac{64}{9}   
\right )
 + 4f(x) 
-\frac{8N_\mu}{3}\ln \frac{m_\mu^2}{m_e^2}
\right ] L_{\rm soft}
- \frac{93}{8} - \frac{5 \pi^2}{2} + 6\zeta(3) - 6\Li_2(x) 
\nonumber \\
&& 
+ 6 \Li_2(1-x) 
+ 3 f(x) 
+N_\mu \left ( 
-\frac{1}{3} \ln^2 \frac{m_\mu^2}{m_e^2} + \left (
\frac{2}{3} \ln \frac{x}{1-x} - \frac{37}{9}
\right )\ln \frac{m_\mu^2}{m_e^2}
\right )
\nonumber \\
&& 
- N_f \left ( 
\frac{8}{3} \Li_2(x) + \frac{281}{27} 
\right ) + N_f g(x)
+N_f^2 \left ( 
-\frac{10}{9} + \frac{(2-x)}{3(1-x+x^2)} \ln(x)
\right ) -\frac{2}{3} N_f N_\mu \ln \frac{m_\mu^2}{m_e^2}\,,
\nonumber 
\ea
and the function $g(x)$ reads
\ba
&& g(x) = (1-x+x^2)^{-2} \left \{
\left (\frac{2}{3}x^4-\frac{5}{4}x^2-\frac{1}{12}x^3
+\frac{17}{12}x-\frac{1}{3} 
\right )\ln^2(x) 
\right.
\nonumber \\
&& \left.
+\left ((1-2x) \left ( \frac{2}{3}x^3-\frac{1}{2}x^2+\frac{2}{3} \right 
)\ln(1-x)
+\frac{37}{9} - \frac{56x}{9} + \frac{47x^2}{6} - 
\frac{67x^3}{18} + \frac{10x^4}{9}
 \right )\ln(x)
\right.
\nonumber \\
&& \left.
-\left (\frac{2}{3}x^2-\frac{7}{6}x+\frac{2}{3} \right )(x^2-x+1)\ln^2(1-x)
+\left (-\frac{10}{3}x^2+\frac{31}{18}x^3+\frac{31}{18}x
-\frac{10}{9}-\frac{10}{9}x^4 \right )\ln(1-x)
\right.
\nonumber \\
&& \left.
+\left (
\frac{11}{12}x^2+\frac{8}{9}x^4-\frac{1}{9}+\frac{2}{9}x-\frac{23}{18}x^3 \right )\pi^2
\right \}.
\ea
Except for the muon contributions, the terms $\delta_2^{(2)}$ and $\delta_2^{(1)}$ were computed in Refs.~ \cite{kur,glov,kur1} and the term $\delta_2^{(0)}$ in Eq.~(\ref{eq6}) was computed in Refs.~\cite{penin,Bonciani:2004gi, Bonciani:2004qt,fer}. We present the result of our computation of the term $\delta_2^{(0)}$ below.

\section{Mass factorization}
\label{mass}

We begin with the description of the method that we use to compute the 
NNLO QED corrections to the Bhabha scattering cross-section. The key to our approach  is a factorization formula that relates massless and massive amplitudes for a given process; such a relationship between massless and massive amplitudes can be expected  
because of the well-known fact \cite{taylor} that in physical gauges collinear 
divergences factorize into the wave function renormalization constants.

We will explain the factorization theorem using the Dirac form factor and then apply it to Bhabha scattering but we stress that the same relation applies to arbitrary scattering amplitudes in QED and QCD in the limit where all
particle masses are much smaller than typical  momentum transfers.
The Soft-Collinear Effective Theory (SCET) \cite{Bauer:2000yr,Bauer:2001yt,Beneke:2002ph} is an appropriate framework to analyze factorization properties
of processes in this limit. This effective field theory is constructed 
by studying the perturbative expansion in QED or QCD and identifying 
those momentum regions in loop integrals 
that lead to singularities once  
the expansion of diagrams in small kinematic variables or masses 
is performed.  These momentum modes 
are described by effective theory fields while the remaining contributions are 
integrated out and absorbed into the Wilson coefficients of the effective theory operators. The singularities relevant to our case arise when particles are soft or have momenta collinear to the external momenta. The effective 
theory description of the electric current requires two different 
collinear fields which interact via soft exchanges.
To explain the structure 
of the result, we  first consider the electron Dirac 
form factor in the limit where the electron momenta fulfill 
$Q^2=-(p_1-p_2)^2\gg p_1^2\sim p_2^2\sim m_e^2$. At  
leading power in the effective theory, 
the vector current $V_\mu = \bar \psi \gamma_\mu \psi$ takes the form
\begin{equation}
V_\mu 
  = 
\int\! ds dt\, {\tilde C}_V(s,t)\, \big[\bar \xi_{2} W_2\big] ( s \bar n_2)\, \gamma_\mu\, \big[W_1^\dagger \xi_{1} \big]( t \bar n_1) \,
+{\cal O}(Q^{-1}).
\end{equation}
Here, $n_1$  and $n_2$ are the light-like reference vectors in the directions of $p_1$ and $p_2$, respectively. The conjugate vectors $\bar n_1$  and $\bar n_2$ point in the opposite directions and fulfill $\bar n_1\cdot n_1=\bar n_2\cdot n_2=2$. These reference vectors must be chosen such that $\bar n_1\cdot p_1 \bar n_2\cdot p_2 = Q^2+{\cal  O}(m_e^2)$. The collinear electron fields $\xi_{1}$ and $\xi_{2}$  are multiplied by light-like collinear Wilson lines 
\begin{equation}\label{eq:Whc}
   W_{i}(x) = \exp\left(ie\int \limits_{-\infty}^0\!ds\,
   \bar{n}_i\cdot A_{c,i}(x+s\bar{n}_i) \right) \,.
\end{equation}
The Fourier transform of the Wilson coefficient
\begin{equation}
C_V(Q^2)\equiv C_V(\bar n_1\cdot p_1 \bar n_2\cdot p_2)\\=
 \int\! ds dt\, {\tilde C}_V(s,t) e^{i s \bar n_2 p_2 - i t \bar n_1 p_1}
\end{equation}
depends only on the hard scale $Q^2$, but is independent of the electron mass. To obtain $C_V(Q^2)$, we perform 
a matching calculation. The simplest way to do the matching is to use dimensional regularization and to calculate the 
on-shell  form factor in the massless theory. In this case all 
loop diagrams in  effective theory vanish and the bare Wilson coefficient 
$C_V(Q^2)$ equals  the on-shell Dirac form factor $\tilde F(Q^2)$ of a massless electron. The massless on-shell form factor has infrared divergences, which show up as poles in $4-d=2\epsilon$. These poles 
correspond to ultra-violet divergences in the effective theory.  
Since the Wilson coefficients are independent of the small electron mass, the difference between  massive and massless 
amplitudes can only arise from matrix elements of  operators in the effective 
field theory. Off-shell Green's functions in the effective theory get contributions from soft interactions between  external legs and collinear interactions in each sector. At  leading power, soft photons have eikonal interactions with collinear fields; only the $n_i\cdot A_s$ component of the soft photon field interacts with collinear electrons moving  in the $i$th direction. These interactions can be removed by field redefinitions \cite{Bauer:2001yt}
\be
\xi_i(x) =S_i(x_-) \xi_i^{(0)}(x)\,,\;\;\;\;A_i^\mu(x) = S_i(x_-) \,{A^{(0)\mu}_i}\, S_{i}^\dagger(x_-)\,,
\label{frdef}
\ee
where $x_-=\frac{1}{2}(\bar{n}_i\cdot x) n_i^\mu$ and the soft Wilson line reads
\begin{equation}\label{eq:Soft}
   S_{i}(x) = \exp\left(ie\int \limits_{-\infty}^0\!ds\,
   n_i\cdot A_{c,i}(x+s n_i) \right) \,.
\end{equation}
The current operator  takes the form
\begin{equation}
V_\mu = 
 \int\! ds dt\, {\tilde C}_V(s,t)\; \big[\bar \xi^{(0)}_{2} W^{(0)}_2\big] ( s \bar n_2)\, \gamma_\mu\,  S^\dagger_2(0) S_1(0) \big[W^{(0)\dagger}_1 \xi^{(0)}_{1} \big]( t \bar n_1) \,
+{\cal O}(Q^{-1}).
\end{equation}
After the field redefinition Eq.~(\ref{frdef}), there is no interaction between the different sectors of the theory. The matrix elements of the 
current  operator factorize into collinear matrix elements for each direction, called  jet-functions, and a soft function, which is given by the matrix element of the soft Wilson lines. This factorization of Green's functions into hard-, jet- and soft functions at large momentum transfers is a well known property of gauge theories \cite{Akhoury:1978vq, Sterman:1978bi, Sen:1982bt}.

For on-shell matrix elements the situation 
is especially simple because  
in the massless 
case  jet and soft functions are trivial since effective theory loop diagrams 
are scaleless. For massive electrons, the collinear matrix elements are functions of the lepton masses, while the soft function also depends on the hard momentum $Q$. We therefore write 
\be
F(Q^2,\{m^2\}) = Z_J(\{m^2\}) S(Q^2, \{m^2  \}) {\tilde F}(Q^2)
+{\cal O}(m^2/Q^2)\,.
\label{Nffac}
\ee 
The relation between the massive form factor $F$ and the massless
form factor $\tilde F$ simplifies when only photonic corrections are
considered. In that case higher order corrections to the soft matrix
element in the massive theory are given by scaleless
integrals and therefore vanish.  This can be seen diagrammatically before performing the somewhat formal decoupling transformation in Eq.~(\ref{frdef}). Hence, we conclude that for photonic corrections the soft
function in Eq.(\ref{Nffac}) is equal to one to all orders in QED
perturbation theory and can be dropped from the right hand side of
Eq.(\ref{Nffac}). The relation between massless and massive form
factors obtained in this way coincides with the relation discussed
recently in Ref.~\cite{mitov}. It is also consistent with the one-loop relation obtained in Ref.~\cite{Catani:2000ef}.

However, a non-trivial soft function appears once
vacuum polarization diagrams with massive particles are considered.
In that case, it is easy to see that the soft momentum 
contribution in the massive theory is not a scaleless integral 
and therefore does not vanish. Moreover, it exhibits non-trivial 
dependence on the hard scale $Q$.   We write the soft 
 matrix element $S(Q^2,\{m^2 \})$ as
\be
S(Q^2,\{m^2\}) = 1 + \sum \limits_{i=e,\mu}^{} \delta S(Q^2,m_i^2)\,,
\ee
where
\be
\delta S(Q^2,m_i^2,N_i) = 
- N_i (4\pi \alpha_0)^2 \int \frac{{\rm d}^d k}{(2\pi)^d}
\frac{p_1 \cdot p_2}{( p_1 \cdot k )\; (p_2 \cdot k) \;k^2}\;i\Pi(k^2,m_i^2). 
\label{sfac}
\ee
In Eq.~(\ref{sfac}) 
$\alpha_0$ stands for  the bare QED coupling constant. 
The vacuum polarization
function $\Pi(k^2)$ in Eq.~(\ref{sfac}) is defined as
\be
i\Pi(k^2,m_i^2) =\frac{(-1)}{(d-1)k^2}
 \int \frac{{\rm d}^dl}{(2\pi)^d}\; {\rm Tr} \left [
\gamma_\alpha \frac{1}{l\hspace{-4.1pt}/ -m_i } \gamma^\alpha \frac{1}{l\hspace{-4.1pt}/ + k\hspace{-5pt}/ -m_i}
\right ].
\ee
The easiest way to compute the soft matrix element is to employ a dispersive representation for the vacuum polarization function $\Pi(k^2)$. Evaluating the integral in Eq.~(\ref{sfac}) in $d=4-2\epsilon$ dimensions, we obtain
\be
\delta S(Q^2,m_i^2,N_i) = 
 N_i a_0^2
m_i^{-4\e} \ln \left ( \frac{Q^2}{m_e^2} \right )\;
\left ( -\frac{1}{12\e^2} + \frac{5}{36\e} - \frac{7}{27}
-\frac{\pi^2}{72}
+{\cal O}(\e)
\right ),
\label{sME}
\ee
where $a_0 = \alpha_0/\pi\; e^{-\gamma \e}(4\pi)^{\e}$ and 
$\gamma$ is the Euler constant. Eq.~(\ref{Nffac}) provides a relation between 
closed lepton loop 
contributions to  massive and massless form factors 
and
allows a derivation  of $N_{e,\mu}$-dependent ${\cal O}(\alpha^2)$ 
corrections to Bhabha scattering 
from the massless results of Ref.~\cite{dixon}.

To determine the square of the jet-function $Z_J$ 
to NNLO in QED, we use Eq.~(\ref{Nffac}) 
and  divide the ratio of 
dimensionally regularized massive and 
massless form factors by the soft matrix element Eq.~(\ref{sfac}).
The expansion of the
massive vector form factor in the limit $Q^2\gg m_e^2$ through 
${\cal O}(\alpha^2)$
can be found in 
Refs.~\cite{gehr1,hoang,kniehl}\footnote{We need the $O(\e^2)$ terms in the 
one-loop contribution to the massive 
form factor which are not provided in Ref.~\cite{gehr1}. 
However, it is simple to compute these terms and
the corresponding result is given in Appendix A. Note also that with the normalization adopted in Eq.~(24) of Ref.~\cite{gehr1} the $\overline {\rm MS}$ result for the contribution of massive quark vacuum polarization to the heavy quark form factor coincides with the result in the on-shell scheme.}.
The massless result can be obtained from 
Refs.~\cite{Moch:2005id,gehr2}.
We express the jet function through the bare QED 
coupling constant $\alpha_0$.
We find
\ba
&& Z_J  = 1 + a_0 m_e^{-2\e}
\left [ \frac{1}{2\e^2} + \frac{1}{4\e} + \frac{\pi^2}{24} +1
+ \e \left ( 2 + \frac{\pi^2}{48} -\frac{\zeta(3)}{6} \right )
+ \e^2 \left ( 4 - \frac{\zeta(3)}{12} + \frac{\pi^4}{320}
+ \frac{\pi^2}{12} \right ) \right ]
\nonumber \\
&& + a_0^2 m_e^{-4\e}
\left [ \frac{1}{8\e^4} + \frac{1}{\e^3} \left ( \frac{1}{8} - \frac{N_f}{24}
\right )
+ \frac{1}{\e^2} \left ( \frac{17}{32} + \frac{\pi^2}{48} -\frac{N_f}{36} 
\right )
+ \frac{1}{\e} \left ( \frac{83}{64} - \frac{\pi^2}{24} + \frac{2\zeta(3)}{3}
\right. \right.
\nonumber \\
&& \left. \left.
-N_f \left ( \frac{209}{432} + \frac{5 \pi^2}{144} \right )
-\frac{N_\mu}{6} \ln \frac{m_\mu^2}{m_e^2}
\right )
 + \frac{561}{128} + \frac{61\pi^2}{192}- \frac{11}{24}\zeta(3)
 - \frac{\pi^2}{2} \ln(2) - \frac{77\pi^4}{2880}
\right.
\nonumber \\
&& \left. 
+N_f \left ( 
\frac{3379}{2592} - \frac{19\pi^2}{216} + \frac{\zeta(3)}{36}
\right )
+ N_\mu \left ( 
\frac{1}{36} \ln^3 \frac{m_\mu^2}{m_e^2}
+ \frac{25}{72} \ln^2 \frac{m_\mu^2}{m_e^2}
+ \left (  
\frac{193}{216} + \frac{\pi^2}{18} 
\right ) \ln \frac{m_\mu^2}{m_e^2}
\right. \right.
\nonumber \\
&& \left. \left.
-\frac{1241}{1296} + \frac{7\pi^2}{54} - \frac{\zeta(3)}{3}
\right )
\right ]
+ {\cal O}\left (\alpha \e^3,\alpha^2\e \right).
\label{eq7.5}
\ea
Setting  $N_f=N_\mu=0$, we find the agreement with  the result of 
Ref.~\cite{mitov}.
The  independence of the jet function $Z_J$ of the hard scale 
$Q^2$ is an explicit demonstration of the mass factorization to 
two-loop order.

Before turning to Bhabha scattering, we want to address a subtlety concerning Eq.~(\ref{Nffac}). This relation relies on the assumption that only hard, collinear and soft momentum modes are relevant in the effective theory computation. However, as was explicitly shown in Ref.~\cite{Smirnov:1999bz}, this assumption is invalid for some diagrams that contribute to the form factor. A particular 
example discussed in Ref.~\cite{Smirnov:1999bz} is the contribution of 
so-called ultra-collinear modes to a two-loop planar vertex 
diagram. Let us stress that the relevance of these modes for 
the full form factor computation would invalidate the factorization formula Eq.~(\ref{Nffac}) since this mode  induces 
additional dependence on the hard scale $Q$ that is not associated 
 with the hard or soft region. It is therefore gratifying to observe 
that the ultra-collinear modes discussed in Ref.~\cite{Smirnov:1999bz}
are not relevant for  the  form factor since their contributions 
cancel out. For example, at the two-loop 
level, the ultra-collinear contribution to a planar vertex diagram cancels exactly against a similar contribution to the non-planar vertex diagram making the full form factor independent of it.

\section{Bhabha scattering} 
\label{comp}

We can now  apply the factorization formula, established for the 
electron Dirac
form factor in the previous Section, to Bhabha scattering.
To get the scattering amplitude ${\cal M}$ in which the electron 
mass is used as a regulator of the collinear singularities, we
only need to multiply the massless amplitude 
$\tilde {\cal M}$ by the square root of the jet function 
$Z_J^{1/2}$
for each electron and positron leg and by the product of soft 
functions that account for soft exchanges in $s,t$ and $u$ channels.
To obtain the soft matrix element, we may use Eq.~(\ref{sME}). While Eq.~(\ref{sME}) is relevant for the space-like electron form factor, we need to generalize it to describe soft exchanges in the $s$- and $u$-channel. We write  
\be
{\cal M}(\{p_i \},\{m^2\}) = Z_J^2(\{m^2\}) \tilde {\cal M}(\{p_i\})
S(s,t,u)
+ {\cal O}({m^2}/{Q^2}),
\label{eq7}
\ee
where the soft function $S(s,t,u)$ is given by 
\be
S(s,t,u) = \left (1 + 2 \sum \limits_{Q^2} 
\sum \limits_{i=e,\mu}^{}
\delta S(Q^2,m_i^2,\lambda_{Q^2} N_i)
\right)
\label{eq23}
\ee
and the sum goes over $Q^2 = -s, -t, -u$ with $\lambda_s = \lambda_t = 1$
and $\lambda_u = -1$. The change $N_i \to -N_i$ shown in Eq.~(\ref{eq23}) 
accounts for the required change in the overall sign in $\delta S$ 
in the $u$-channel.

We can now use the two-loop result for the Bhabha 
scattering amplitude computed in the massless limit \cite{dixon} and 
employ  Eqs.~(\ref{eq7.5}), (\ref{eq7}) and (\ref{eq23}) 
to obtain the scattering amplitude in which  collinear 
divergences are regularized  
by the electron mass and infrared divergences 
are regularized dimensionally.  
With the massive  scattering amplitude at hand, the computation 
of the two-loop QED corrections to the large-angle Bhabha scattering 
cross-section becomes straightforward; what we need in addition is 
the cross-sections for the inelastic processes $e^+e^- \to e^+ e^- +n\, \gamma$, with $n=1$ and $n=2$,
in the soft photon approximation.
We write the perturbative 
expansion of the Bhabha scattering cross-section as 
\ba
&& \frac{{\rm d}\sigma}{{\rm d} \Omega} = 
\exp{(\displaystyle \frac{\alpha}{\pi} F_{\rm soft})}\:Z_J^4 \; |S|^2\;\: \frac{\bar \alpha^2}{s}
\left (  {\rm d} \sigma_{0} + \frac{\bar \alpha}{\pi} {\rm d} \sigma_1^{v}
+ \left ( \frac{ \bar \alpha}{\pi} \right )^2 
\left ( {\rm d} \sigma_{1\times1}^{v}  + {\rm d}\sigma_2^{v} \right )
+{\cal O}(\alpha^3)
\right ).
\label{eq8}
\ea
In this formula, $Z_J$ is the square of the jet-function  given 
in Eq.~(\ref{eq7.5}); $F_{\rm soft}$ describes soft photon radiation
which, in case of QED, is known to factorize and exponentiate. Up to an overall factor $\bar\alpha^2/s$ the quantity ${\rm d}\sigma_0$ is the tree level cross section in $d$-dimensions and ${\rm d}\sigma^v_1$ denotes the one-loop virtual contributions. At two loops, there are two types of virtual corrections: the quantity ${\rm d}\sigma_2^{v}$ contains the interference of the two-loop amplitude with the tree level amplitude, while ${\rm d}\sigma_{1\times 1}^{v}$ describes the interference of one-loop amplitude with itself. These contributions are to be computed with massless leptons. The massive result for the virtual corrections is then obtained by multiplying the massless result with $Z_J^4\, |S^2|$. The  massless cross-sections ${\rm d} \sigma_1^{v}$ and ${\rm d}\sigma_2^{v}$  
can be found in Ref.~\cite{dixon}, while ${\rm d}\sigma_{1\times 1}^{v}$ can be obtained from Ref.~\cite{babis}, as we explain below. 
Note also that results of  Refs.~\cite{dixon,babis} are written 
through the QED coupling constant $\bar \alpha$ 
renormalized in the ${\overline {\rm MS}}$ scheme. It is for this 
reason that $\bar \alpha$ appears in Eq.~(\ref{eq8}).

In Ref.~\cite{babis} the interference of the 
one-loop amplitude with itself 
was  obtained for  quark-quark scattering in QCD. 
To extract the QED piece relevant for Bhabha scattering  
from these results, we need to analyze the 
color algebra; such an analysis shows that ${\rm d}\sigma_{1\times 1}^{v}$
can be obtained from the  computation of Ref.~\cite{babis} by 
taking the $N \to 0$ limit of the  QCD result, where $N$ is the number of 
colors,  
and subtracting from 
it suitably weighted products of the 
one-loop Bhabha scattering cross-section ${\rm d}\sigma_1^{v}$ and 
the electron Dirac form factor in the massless approximation.
Note that  the divergent terms in ${\rm d}\sigma^{v}_{1\times1}$ 
can be obtained from Catani's decomposition \cite{catani} of 
the one-loop scattering amplitude for $e^+e^- \to e^+e^-$ and 
the ${\cal O}(\alpha)$ correction to the Bhabha scattering 
cross-section in dimensional regularization derived in Ref.~\cite{dixon}. 
For this  reason we only present the finite part of ${\rm d}\sigma^{v}_{1\times1}$; 
it is given in the Appendix B.

Finally, the description of soft radiation in Eq.~(\ref{eq8}) is encapsulated in the function $F_{\rm soft}$; we require 
this function for massive electron-positron scattering. Since $F_{\rm soft}$ describes the emission of a real photon, it is simplest to evaluate it in the on-shell scheme. In doing so, the effects of vacuum polarization contributions are absorbed into the coupling constant and do not 
need to be evaluated explicitly. 
The function $F_{\rm soft}$ is determined by the  integral 
\be
F_{\rm soft} = 
-4\pi^2\;\int \limits_{k_0 \le \omega_{\rm cut}}^{} 
\frac{{\rm d}^d k}{(2\pi)^{d-1} 2 k_0} \;\; J_\mu J^\mu,
\label{eq9}
\ee
where the soft current $J^\mu$ is given by
\be
J^\mu = \sum \limits_{i}^{} q_i \lambda_i\;\frac{p_i^\mu}{p_i\cdot k},
\ee 
$q_i$ is the charge of the particle $i$ in units of the positron 
charge and $\lambda = \pm 1$ for the incoming(outgoing) particle, 
respectively. Writing 
\be
J_\mu J^\mu = 
\sum \limits_{i \ne j} 
\frac{p_i\cdot p_j}{(p_i\cdot k)(p_j\cdot k)}q_iq_j \lambda_i \lambda_j 
+ \sum \limits_{i}^{} \frac{m_e^2}{(p_i\cdot k)^2},
\ee
we observe that two types of integrals are required for the evaluation 
of $F_{\rm soft}$. We give expressions 
for these integrals below neglecting all the terms that are suppressed 
by powers of the electron mass.

The first integral depends on the relative 
momenta of two charged particles. We find
\ba
&& I_{ij} = 
\int  \limits_{k_0 \le \omega_{\rm cut}}^{}
\frac{{\rm d}^{d-1} k}{(2\pi)^{d-1} 2k_0}\;\;\frac{p_i\cdot p_j}{(p_i\cdot k)(p_j\cdot k)} 
=-{\cal N}_\e 
\frac{(2\omega_{\rm cut})^{-2\e}}{2\e}
\left [ \ln \frac{1}{1+x_{ij}} 
-L_m 
\right.
\nonumber \\
&& \left. + \e \left ( \frac{1}{2} L_m^2 
+\Li_2 \left ( -x_{ij}\right ) + \frac{\pi^2}{3} \right )
 + \e^2 \left ( -\frac{L_m^3}{6} - \frac{\pi^2}{3}L_m
+ 2\zeta(3) + \Li_3 \left ( -x_{ij} 
\right)
\right )
\right ],
\ea
where $x_{ij} = (1+\cos \theta_{ij})/(1-\cos \theta_{ij})$, 
$\theta_{ij}$ is the relative angle between the three-momenta 
of the particles $i$ and $j$ and the normalization factor ${\cal N}_\e$ 
reads
\be
{\cal N}_\e = \frac{\Gamma(1-\e)}{4\pi^2 (4\pi)^{-\epsilon} \Gamma(1-2\e)}.
\ee
Also, we introduced $L_m = \ln(m_e^2/s)$ 
to denote the collinear logarithm.The second integral required to describe the 
soft radiation reads
\ba
&& I_{s} =
\int  \limits_{k_0 \le \omega_{\rm cut}}^{}
\frac{{\rm d}^{d-1} k}{(2\pi)^{d-1} 2k_0}\;\;\frac{m_e^2}{(p_i\cdot k)^2}  
=-{\cal N}_\e
\frac{(2\omega_{\rm cut})^{-2\e}}{2\e}
\left [ 1-\e \ln \frac{m_e^2}{s} 
+\frac{\e^2}{2}\ln^2 \frac{m_e^2}{s} \right ],
\ea
where we used the on-shell condition $p_i^2 = m_e^2$.

We now have everything in place to calculate the NNLO QED corrections 
to Bhabha scattering. We substitute all the necessary ingredients 
into Eq.~(\ref{eq8}). To combine the different pieces, it is simplest to first express the on-shell coupling constant appearing in the soft radiation exponential in Eq.~(\ref{eq8}) through the ${\overline {\rm MS}}$ coupling constant. In $d$-dimensions, the relevant relation reads
\be
\alpha=\bar\alpha(\mu)\, \mu^{2\epsilon}\left\{1-\frac{2\bar\alpha(\mu)}{3\pi}\sum_{f=e,\mu} \left[\ln\frac{\mu}{m_f}+\epsilon\left(\ln^2\frac{\mu}{m_f}+\frac{\pi^2}{24}\right)+{\cal O}(\epsilon^2)\right]\right\}\,.
\ee
After adding the real and virtual parts, all infrared divergences cancel and we can use the four-dimensional relation between the ${\overline {\rm MS}}$ and the on-shell fine structure constants 
\ba
\bar \alpha(\mu) &=& 
\alpha \left \{ 
1 + \left ( \frac{\alpha}{\pi} \right) 
\left ( \frac{2N_f}{3}\ln \frac{\mu}{m_e}
-\frac{2N_\mu}{3}\ln \frac{m_\mu}{m_e} 
\right )
+ \left ( \frac{\alpha}{\pi} \right )^2 
\left [ 
\frac{4N_f^2}{9} \ln^2 \frac{\mu}{m_e} + \frac{N_f}{2} \ln \frac{\mu}{m_e} 
\right. \right.
\nonumber \\
&& \left. \left.
+ \frac{15N_f}{16}
-\frac{8N_f N_\mu}{9} \ln \frac{\mu}{m_e} \ln \frac{m_\mu}{m_e}
-\frac{N_\mu}{2} \ln \frac{m_\mu}{m_e} 
+\frac{4N_\mu^2}{9} \ln^2 \frac{m_\mu}{m_e} 
\right ]
\right \},
\ea
 to obtain the expansion of the Bhabha scattering cross section through  NNLO in terms of the on-shell QED coupling constant. Upon doing so, we reproduce the formulas for radiative corrections to Bhabha scattering shown in Section~\ref{not} and obtain an explicit expression for $\delta_2^{(0)}$, which is presented below.
We write  $\delta_2^{(0)}$ in the following way
\be
\delta_2^{(0)} = \delta_2^{(0,1)} + N_f \delta_2^{(0,2)}
+ N_\mu \delta_2^{(0,3)}
+N_f^2 \delta_2^{(0,4)}
+N_f N_\mu \delta_2^{(0,5)}
+\frac{N_\mu^2}{3} \ln^2 \frac{m_\mu^2}{m_e^2}
,
\ee
where we separate photonic corrections and the corrections caused 
by closed lepton loops. We now present the results for these 
terms separately. For the photonic corrections we find
\ba
\delta_2^{(0,1)} &=& 8 {\cal L}_{\rm soft}^2 + (1-x+x^2)^{-2}  {\cal L}_{\rm soft}\Big(-x^2(4x^2+5-6x)\pi^2
-x(3-3x^2-x+4x^3)\ln^2(x) \nonumber\\[2mm]
&&+\left [ 2x^2(4x^2+5-6x)\ln(1-x)
-12+16x-18x^2+6x^3 \right ]\ln(x)
+2x(x^2+1)\ln(1-x) \nonumber\\[2mm]
&& +2(x^2-x+1)(2x^2-3x+2)\ln^2(1-x)
+16(x^2-x+1)^2 \left ( 1 + \Li_2(x)  \right ))\Big) \nonumber\\
&&+8 \Li_2(x)^2 + \frac{27}{2}- 2\pi^2 \ln(2) 
+ (1-x+x^2)^{-2}\Delta_{2}^{(0,1)}\,, 
\ea
where ${\cal L}_{\rm soft} = \left ( 1 - \ln(x/(1-x) \right) L_{\rm soft}$,
and 
\ba
&& \Delta_{2}^{(0,1)} = 
+\left( \frac{31}{480}x^4-\frac{8}{45}+\frac{37}{90}x^2-\frac{7}{72}x
-\frac{47}{180}x^3 \right )\pi^4
+\left [
\frac{1}{48}x(35x^3-2x^2+20x+24)\ln^2(x)
\right.
\nonumber \\
&& \left. 
+\left ( \left ( -\frac{35}{24}x^4+\frac{8}{3}x^3-\frac{11}{12}x^2
-\frac{5}{2}x+\frac{8}{3} \right )\ln(1-x)
-\frac{15}{8}x^3+\frac{11}{12}x^2+\frac{23}{12}x-\frac{1}{2} \right )\ln(x)
\right.
\nonumber \\
&& \left.
+\left (-\frac{5}{48}x^4+\frac{1}{12}x^3-\frac{7}{3}x^2+3x
-\frac{49}{24} \right )\ln^2(1-x)
+\frac{1}{24}x(43x^2-74x+24)\ln(1-x)
\right.
\nonumber \\
&& \left.
-\frac{3}{4}x^2+\frac{17}{8}+\frac{83}{24}x^3
-\frac{19}{8}x^4-\frac{61}{24}x \right ]\pi^2
+\frac{1}{96}(43x^3-8x^2+5x+14)x\ln^4(x)
\nonumber \\
&& + \left (  \left (-\frac{43}{24}x^4+\frac{7}{6}x^3
+\frac{1}{2}x^2-\frac{17}{12}x+ \frac{2}{3} \right )\ln(1-x)
-\frac{1}{24} (16x^2+30x-67)x\right )\ln^3(x)
\nonumber \\
&& 
+\left ( \left (\frac{19}{16}x^4-\frac{29}{8}x^3+\frac{7}{8}x^2
+\frac{39}{8}x-\frac{9}{2} \right )\ln^2(1-x)
+\frac{1}{8}(9x^2-2x-24)x\ln(1-x)
\right.
\nonumber \\
&& \left.
-\frac{9}{2}x^4+\frac{29}{8}x^3
+\frac{17}{8}x^2-\frac{43}{8}x+\frac{9}{2} 
\right )\ln^2(x)
+\left (
\left (-\frac{1}{8}x^4+\frac{1}{3}x^3+\frac{11}{3}x^2-\frac{37}{6}x
+4 \right )\ln^3(1-x)
\right. 
\nonumber 
\ea\ba
&&  \left. + \left ( \frac{13}{8}x^3-\frac{11}{2}x^2+\frac{27}{4}x-3 \right )
\ln^2(1-x)
+\frac{1}{4}x(51x-22+36x^3-64x^2)\ln(1-x)
\right.
\nonumber \\
&& \left.
+(12-12x+8x^2-x^3)\zeta(3)
-\frac{279}{16}x^2+\frac{231}{16}x+\frac{93}{16}x^3-\frac{93}{8} 
\right )\ln(x)
\nonumber
 \\
&& 
+ \left ( \frac{1}{32}-\frac{3}{4}x+\frac{71}{48}x^2-\frac{29}{24}x^3
+\frac{9}{32}x^4 \right )\ln^4(1-x)
+\frac{1}{24}(9x^2+4x+9)x\ln^3(1-x)
\nonumber 
\\
&& +\left (\frac{45}{4}x^2-6x+\frac{7}{2}x^4-6x^3+\frac{7}{2} 
\right)\ln^2(1-x)
+\left ( (-4x^3-6+6x-x^2)\zeta(3)+3x(x^2+1) \right )\ln(1-x)
\nonumber \\
&& +\left (-9+\frac{43}{2}x-34x^2+22x^3-9x^4 \right )\zeta(3)
+ \left [ 
\left (-\frac{17}{4}x-3x^4+2+\frac{9}{4}x^2
+\frac{7}{4}x^3 \right )\ln^2(x)
\right.
\nonumber \\
&&  \left.
+\left ( \left (-6+7x+\frac{15}{2}x^2+8x^4-14x^3 \right )\ln(1-x) + 
+\frac{27}{2}x-13x^2+4x^3-12 \right )\ln(x)
\right. 
\nonumber 
\\
&&  \left. 
+\frac{1}{2}\left (x^2-4x+7 \right )\left (2x^2-3x+2 \right )\ln^2(1-x)
+\frac{1}{2}x(5+3x^2)\ln(1-x) 
\right.
\nonumber 
\\
&& \left. -32x-32x^3+16x^4+16+48x^2
+\pi^2 \left( \frac{31}{6}x^3+\frac{5}{3}(1-x)
-\frac{15}{4}x^2-\frac{8}{3}x^4 \right )
 \right ]\Li_2(x)
\nonumber \\
&&+\left ((-6+5x+3x^2-5x^3)\ln(x)
+2(1-x^2)(3x^2-5x+3)\ln(1-x)
+\frac{x}{2}(1-x^2) \right )\Li_3(1-x)
\nonumber \\
&& +\left ((-4-x+x^2+2x^3-2x^4)\ln(x)
+(x^2+6+4x^3-6x)\ln(1-x)+\frac{x}{2}(4x^2-10x+5) \right )\Li_3(x)
\nonumber \\
&& +\left (-6+4x+\frac{9}{2}x^2-7x^3 \right )\Li_4\left (\frac{x}{x-1} \right )
+\frac{x}{2}\left (12x^3+14-9x-8x^2 \right )\Li_4(1-x)
\nonumber \\
&& -\frac{1}{2}(1-x^2)(4x-1)(x-4)\Li_4(x).
\ea

The contribution of diagrams with a single electron or muon  
vacuum polarization insertion is described by $\delta_2^{(0,2)},\;\delta_2^{(0,3)}$. We obtain
\be
\delta_2^{(0,2)} =
{\cal L}_{\rm soft} \left ( \frac{40}{9} + \frac{4(x-2)}{3(1-x+x^2)} \ln(x) 
\right )
+ \frac{40}{9} \Li_2(x)  
+\frac{1967}{108}  
+(1-x+x^2)^{-2}\Delta_2^{(0,2)},
\ee
where 
\ba
&& \Delta_2^{(0,2)} =
\left ( \frac{2}{3}x^2-x+\frac{4}{3} \right )(x^2-x+1)\Li_3(x)
+\frac{2}{3}(1-x^2)(x^2-x+1)\Li_3(1-x)
\nonumber \\
&& 
+\left ( - (x^2-x+1) \left (\frac{2}{3}x^2-\frac{7}{3}x+4 \right )\ln(x)
+\frac{2}{3}(x^2-x+1)(1-x^2)\ln(1-x) \right )\Li_2(x)
\nonumber \\
&& 
+ \left(
-\frac{1}{18}x^3-\frac{11}{18}x^2+\frac{1}{9}x^4
+\frac{31}{36}x-\frac{1}{9} \right )\ln^3(x)
+\left (
 \left (- \frac{4}{3}x^2-\frac{1}{3}x^4+\frac{1}{3}x-\frac{1}{3}+x^3
\right )\ln(1-x)
\right.
\nonumber \\
&& \left.
-\frac{10}{9}x^4+\frac{14}{3}x^2-\frac{4}{9}x^3-\frac{46}{9}x+\frac{55}{18}
\right )\ln^2(x)
+ \left (
-x \left ( \frac{7}{12}x-\frac{1}{3}+\frac{x^3}{3}-\frac{x^2}{2}
\right )\ln^2(1-x)
\right.
\nonumber \\
&& \left.
+\left (\frac{1}{2}x^2-\frac{10}{9}-\frac{25}{9}x^3+\frac{37}{18}x
+\frac{20}{9}x^4 \right )\ln(1-x)
+\left (-\frac{1}{36}x^2+\frac{1}{3}x^4-\frac{1}{9}+\frac{2}{3}x-\frac{5}{9}x^3
\right )\pi^2
\right.
\nonumber \\
&& \left.
-\frac{337}{18}x^2+\frac{449}{54}x^3-\frac{281}{27}
+\frac{418}{27}x
-\frac{56}{27}x^4 \right )\ln(x)
-\frac{1}{9}(x^2-x+1)(1-x)^2\ln^3(1-x)
\nonumber \\
&& 
+ \left ( \frac{10}{9}+\frac{9}{2}x^2+\frac{10}{9}x^4-\frac{29}{9}x
-\frac{29}{9}x^3 \right )\ln^2(1-x)
+\left (  \left (- \frac{16}{9}x^2-\frac{2}{9}x^4-\frac{4}{9}+\frac{11}{9}x
+x^3 \right )\pi^2
\right.
\nonumber \\
&& \left.
+ \frac{56}{9}x^2-\frac{161}{54}x^3+\frac{56}{27}-
\frac{161}{54}x+\frac{56}{27}x^4 \right )\ln(1-x)
+\left ( \frac{47}{12}x^3-\frac{7}{3}x^4-\frac{1}{6}
+\frac{11}{12}x-\frac{71}{18}x^2 \right )\pi^2
\nonumber \\
&& +(x^2-x+1)\left ( 2x^2-\frac{5x}{3}+\frac{4}{3} \right )\zeta(3), 
\ea
and
\ba 
&& \delta_2^{(0,3)} =  \frac{8}{3} \ln \frac{m_\mu^2}{m_e^2}\;{\cal L}_{\rm soft}
+\frac{1}{9} \ln^3 \frac{m_\mu^2}{m_e^2}
+ \left ( \frac{19}{18} - \frac{1}{3} \ln \frac{x}{1-x} \right ) 
\ln^2 \frac{m_\mu^2}{m_e^2}
\nonumber \\
&& + \left ( \frac{8}{3} \Li_2(x) + \frac{191}{27} 
 + (1-x + x^2)^{-2}\Delta_2^{(0,3)}
\right) \ln \frac{m_\mu^2}{m_e^2}
+ \frac{14 \pi^2}{27} - \frac{4}{3}\zeta(3) - \frac{1241}{324},
\ea
where 
\ba
&& \Delta_2^{(0,3)} =
(1-x+x^2) \left ( \frac{2}{3}x^2 - \frac{7}{6}x+\frac{2}{3} \right ) \ln^2(1-x)
+\left ( (2x-1) \left ( \frac{2}{3}x^3 - \frac{1}{2}x^2 + \frac{2}{3} 
\right )\ln(x) 
\right.
\nonumber \\
&& \left. 
+ \frac{10}{9} + \frac{10}{9}x^4 
- \frac{31}{18} x 
- \frac{31}{18} x^3 
+ \frac{10}{3}x^2 \right ) \ln(1-x)
+ \left ( \frac{1}{12}x^3 - \frac{2}{3}x^4 - \frac{17}{12}x
+ \frac{5}{4}x^2 + \frac{1}{3} \right ) \ln^2(x)
\nonumber \\
&& + \left ( - \frac{37}{9} - \frac{10}{9} x^4 + \frac{56}{9}x
+ \frac{67}{18}x^3 - \frac{47}{6}x^2 \right ) \ln(x)
+ \left ( \frac{23}{18}x^3 - \frac{8}{9}x^4 + \frac{1}{9} - \frac{11}{12}x^2
- \frac{2}{9}x \right ) \pi^2.
\ea

The corrections with two insertions of the closed lepton  loop read
\be
\delta_2^{(0,4)} = \frac{25}{27} + (1-x+x^2)^{-2} \Delta_{2}^{(0,4)},
\ee
where
\ba
\Delta_2^{(0,4)} && = 
\left ( \frac{1}{3}-\frac{x^3}{9}
+\frac{7x^2}{18}-\frac{4x}{9}
\right )\ln^2(x)
-\frac{5}{9}(x^2-x+1)(2-x)\ln(x)
\nonumber \\
&& 
+x \left ( \frac{1}{9}-\frac{5x}{18}-\frac{x^3}{9}
+\frac{2x^2}{9} \right 
)\pi^2.
\ea
and 
\be
\delta_2^{(0,5)} = 
\left ( \frac{10}{9} + \frac{(x-2)}{3(1-x+x^2)}\ln x \right )\ln \frac{m_\mu^2}{m_e^2}.
\ee
Because the above expressions are rather lengthy, we have included them in electronic form in our submission to the arXiv.

\section{Conclusion}
\label{conc}
We presented a novel relation between massive and massless 
scattering amplitudes in QED valid in the limit 
when all kinematic invariants are large compared to 
masses of particles that participate in the scattering process; for 
quenched QED, our result agrees with a similar relation between massive 
and massless scattering amplitudes discussed recently in \cite{mitov}.
We used this relation to derive the  
NNLO QED corrections for Bhabha scattering  confirming 
earlier results on photonic and electron loop corrections 
of Ref.~\cite{penin,Bonciani:2004gi, Bonciani:2004qt,fer}. We also obtained 
NNLO contributions to Bhabha scattering due to muon and tau 
vacuum polarization loops that were not available in the literature. 

In variance with the approach discussed in Ref.~\cite{penin}, 
our method is directly applicable to QCD.  A potentially interesting 
application is the computation of  the NNLO QCD virtual corrections to 
heavy (e.g.~$b$) quark production at moderate to large momentum transfers 
at the Tevatron and the LHC using the two-loop matrix elements for 
the $gg \to q \bar q$ process computed in Ref.~\cite{babis2} 
for massless quarks.

\vspace*{0.8cm}
{\bf Acknowledgments}  
This research was supported in part by the US Department of
Energy contracts DE-AC02-07CH11359 and DE-FG03-94ER-40833,
the DOE Outstanding Junior Investigator Award 
and by the Alfred P. Sloan Foundation. Fermilab is operated 
by the Fermi Research Alliance, LLC under the contract 
with the US Department of Energy.

\vspace*{0.8cm}
{\bf Note added}  
While we were finalizing our paper, Ref.~\cite{Actis:2007gi} appeared in which the muon-loop contribution to Bhabha scattering was evaluated. After the authors of Ref.~\cite{Actis:2007gi} corrected one of the master integrals their result agrees with ours.

\begin{appendix}

\section{The one-loop massive Dirac form factor}
The one-loop massive Dirac space-like 
form factor, in the limit $Q^2 \gg m_e^2$, reads
\ba
 F_1 && = 1 + \frac{\bar \alpha(m_e)}{2\pi} \frac{\Gamma(1+\e)}{(4\pi)^{-\e}}
\left [ 
\frac{L-1}{\e} -\frac{L^2}{2} + \frac{3L}{2} - 2 + \frac{\pi^2}{6}
+ \e \left ( 
\frac{L^3}{6} - \frac{3L^2}{4} + \left ( 4 - \frac{\pi^2}{6} \right )L
+2 \zeta(3) 
\right. \right.
\nonumber \\
&& \left. \left.
+ \frac{\pi^2}{4}-4 \right )
 + \e^2 \left ( 
-\frac{L^4}{24} + \frac{L^3}{4} + \left ( \frac{\pi^2}{12}-2 \right ) L^2
+ \left( 8 - \frac{\pi^2}{4}-2\zeta(3) \right) L 
\right. \right.
\nonumber \\
&& \left. \left.
+ 3\zeta(3) + \frac{2\pi^2}{3}-8+\frac{\pi^4}{40} 
\right )
\right ], 
\ea
where $L = \ln(Q^2/m_e^2)$ and $\bar \alpha(m)$ is the $\overline {\rm MS}$
fine structure constant evaluated at the scale $m$.

\section{NNLO results}

Here, we give the finite part of the ${\cal O}(\alpha^2)$ correction to the cross-section due to the interference of the {\it photonic} 
one-loop amplitude with itself. The divergent part can be obtained using Catani's results \cite{catani}.
\ba
&& s^{2\e}{\rm d} \sigma^{v}_{1\times 1}|_{\rm finite} = 
\frac{(8x^4-13x^3+13x^2-7x+4)}{2x^2}\Li_4 \left ( \frac{x}{x-1} \right )
\nonumber \\
&& 
-\frac{(8-13x+13x^2-7x^3+4x^4)}{2x^2}\Li_4(1-x)
-\frac{(1-x^2)(2x^2-3x+2)}{x^2}\Li_4(x)
\nonumber \\
&& +\left \{ \frac{(-x^2-9x^3+8x^4+9x-4)}{2x^2}\ln(x)
+\frac{(1-x^2)(2x^2-3x+2)}{x^2}\ln(1-x)
\right.
\nonumber \\
&& \left.
 +\frac{3(1-x^2)(2x^2-3x+2)}{2x^2} \right \}\Li_3(1-x)
+\left \{
-\frac{(13x^2-7x^3+4x^4-13x+8)}{2x^2}\ln(x)
\right.
\nonumber \\
&& \left.
+\frac{(4-3x-x^2+3x^3)}{2x^2}\ln(1-x)
+\frac{(12x^4-9x^3+24-27x+26x^2)}{4x^2}
\right \} \Li_3(x) 
\nonumber \\
&& + \left \{ \frac{3(13x^2-7x^3+4x^4-13x+8)}{4x^2}\ln^2(x)
+ \left ( \frac{(-13x^2+x^3+19x-12)}{2x^2}\ln(1-x)
\right. \right.
\nonumber \\
&& \left. \left. -\frac{(12x^4-9x^3+24-27x+26x^2)}{4x^2} \right )\ln(x)
+\frac{(1-x^2)(2x^2-3x+2)}{2x^2}\ln^2(1-x)
\right.
\nonumber \\
&& \left.
+\frac{3(1-x^2)(2x^2-3x+2)}{2x^2}\ln(1-x)
-\frac{(13x^2-7x+4-13x^3+8x^4) \pi^2}{4x^2}
\right \}\Li_2(x)
\nonumber \\
&& -\frac{(45x^4+26x^3-241x^2+230x-64)}{96x^2}\ln^4(x)
+\left( \frac{(45x^4-38x^3-8x^2+28x+16)}{24x^2}\ln(1-x)
\right.
\nonumber 
\ea
\ba
&& \left.
+\frac{(12x^4+15x^3-93x^2+169x-96)}{24x^2} \right )\ln^3(x)
+ \left(  -\frac{(12x^4-16x^3+13x^2+27x-24)}{8x^2}\ln(1-x)
\right.
\nonumber \\
&& \left.
+\frac{(11x^4-10x^3-38x^2+62x-40)}{16x^2}\ln^2(1-x)
 +\frac{(19x^4+14x^3+70x^2-100x+80)\pi^2}{48x^2}
\right.
\nonumber 
\\
&& \left.
 -\frac{(154x^3-421x^2+460x-328)}{16x^2} \right )\ln^2(x)
 + \left [ -\frac{(27x^4-30x^3-16x^2+54x-32)}{24x^2}\ln^3(1-x)
\right. \nonumber \\
&& \left.
 -\frac{(12x^4-7x^3-29x^2+42x-24)}{8x^2}\ln^2(1-x)
+ \left (-\frac{(35x^4+2x^3-82x^2+86x-48) \pi^2}{24x^2}
\right. \right.
\nonumber 
\\
&& \left. \left.
+\frac{(178x-155x^2+44x^3-128)}{8x^2} \right )\ln(1-x)
 +\frac{(12x^4-100x^3+74x^2-11x+24)\pi^2}{24x^2}
\right.
\nonumber 
\\
&& \left.
+ \frac{(56x^4+165x^2-103x^3-121x+68)}{6x^2}\zeta(3)
-\frac{(599x^2-582x-302x^3+448+128x^4)}{8x^2}
\right ]\ln(x) 
\nonumber  \\
&& +\frac{(3x^4+14x^3-18x^2+26x-5)}{96x^2}\ln(1-x)^4
-\frac{(14x^2+4x^4-9x^3-9x+4)}{8x^2}\ln^3(1-x)
\nonumber 
\\
&& +\left ( -\frac{(5x^4-42x^3+92x^2-78x+26)}{48x^2}\pi^2
-\frac{(-23x+21+21x^2)}{8x} \right )\ln^2(1-x)
\nonumber 
\\
&& 
+\left ( \frac{(23x^3-137x^2+134x-60)}{24x^2}\pi^2
-\frac{(56x^4+165x^2-103x^3-121x+68)}{6x^2}\zeta(3)
\right. \nonumber 
\\
&& \left.
+\frac{(115x^2-76x-76x^3+64+64x^4)}{4x^2} \right )\ln(1-x)
+\frac{(-260+742x-724x^2+34x^3+257x^4)}{1440x^2}\pi^4
\nonumber 
\\
&& -\frac{(229x^2-200x^3+82x^4-98x+82)}{48 x^2}\pi^2
-\frac{(-295x-241x^3+386x^2+168x^4+204)}{12x^2}\zeta(3)
\nonumber \\
&& +\frac{(-464x^3+288x^4+599x^2+288-464x)}{4x^2}.
\ea

\end{appendix}

\end{document}